\documentclass[prl,floatfix,preprint,superscriptaddress]{revtex4}
\usepackage[english]{babel}
\usepackage{amsmath,amsfonts,amssymb,float}
\usepackage{graphicx}

\def\pd#1#2{\frac{\partial #1}{\partial #2}}

\def\RAL{Central Laser Facility, Rutherford Appleton Laboratory, Chilton, Didcot, Oxon OX11 0QX, UK}
\def\IST{GoLP/Instituto de Plasmas e Fus\~ao Nuclear - Laboratorio Associado, Instituto Superior T\'ecnico, 1049-001 Lisbon, Portugal}
\def\Strath{SUPA, Department of Physics, University of Strathclyde, Glasgow, G4 0NG, United Kingdom}
\def\UCLA{Dept. of Physics and Astronomy, University of California Los Angeles, Los Angeles CA 90095-1547, USA}
\def\oxford{Department of Physics, University of Oxford, Oxford OX1 3PU, UK}

\begin{document}

\title{Plasma wakefields driven by intense, broadband, incoherent
  electromagnetic radiation}

\author{R.M.G.M. Trines}
\affiliation\RAL
\author{L.O. Silva}
\affiliation\IST
\author{J.T. Mendon\c{c}a}
\affiliation\IST
\author{W.B. Mori}
\affiliation\UCLA
\author{P.A. Norreys}
\affiliation\RAL
\affiliation\oxford
\author{R. Bingham}
\affiliation\RAL
\affiliation\Strath

\begin{abstract}
Non-linear wave-driven processes in plasmas are normally described by either a
monochromatic pump wave that couples to other monochromatic waves, or as a
random phase wave coupling to other random phase waves. An alternative
approach involves an incoherent, random or broadband pump coupling to
monochromatic and/or
coherent structures in the plasma. This approach can be implemented through
the wave kinetic model. In this model, the incoming pump wave is described by
either a bunch (for coherent waves) or a sea (for random phase waves) of
quasi-particles. A particle-in-cell type code has been developed to perform
numerical simulations of such interactions using the quasi-particle
approach. This code allows for a comparatively easy description of both random
phase and coherent pump pulses coupling to slow electrostatic plasma waves,
while providing an extended range of powerful diagnostics leading to a deeper
physical insight into the dynamics of the fast waves. As an example, the
propagation of short, intense laser pulses through a plasma has been
simulated. A sample of the phenomena that can be studied for this case
includes modulational instabilities, photon wave breaking and turbulence, and
pulse compression, stretching, and chirping. The extensibility of the
numerical implementation to other types of fast wave-slow wave interactions
is also discussed.
\end{abstract}

%\begin{keyword}
%Wave kinetics; Broadband wave-plasma interactions; Numerical methods;
%Modulational Instabilities; Photon acceleration
%\PACS
%\end{keyword}

\maketitle

\section{Introduction}

It is a well-known fact that in the numerical modelling of plasma physics, the
plasma particles can be described as if they were a fluid. Fluid codes, Vlasov
codes, and magneto-hydrodynamic codes are all based on this observation. The
fact that in the interaction of fast, short-wavelength waves with a plasma,
the fast wave modes can actually be described as a bunch of particles is less
known. However, in the case that there is a clear separation in time and
length scales between populations of ``fast'' and ``slow'' waves in the
plasma, it may be worthwile to regard the ensemble of fast modes as a
distribution of quasi-particles. This approach to wave-plasma interaction is
commonly called \emph{wave kinetics}~\cite{sagdeev}. Tappert and
Besieris~\cite{bes73} identified the power of the
Wigner-Moyal~\cite{wig32,moyal49} formalism to study classical wave equations
using phase space distribution functions.  The field is represented by a
Wigner distribution of quasi-particles (photons, plasmons etc), and in the
limit of short wavelengths the transport equation is formally equivalent to
the Vlasov equation.  An analogy between particle dynamics and field dynamics
can then be easily established, leading to new and revealing physical insight.
A similar formalism has been developed in solid state physics to describe
phonon interactions~\cite{peier55}, while the Wigner-Moyal formalism has also
been used to study electromagnetic wave propagation in random media in the
geometrical optics approximation~\cite{mcdon88}. It has been successfully used
in the study of weak turbulence in plasmas since the 1960's~\cite{kadom65} and
more recently in the study of photon Landau damping~\cite{bing97} and
ultra-intense laser-plasma interactions~\cite{silva00}. The Vlasov-like
structure of the photon kinetic equation provides the ideal setting to examine
photon dynamics under the particle-in-cell (PIC)
framework~\cite{bes73,dawson83,bird}. The most powerful features of the
photon kinetic code is its simplicity and the natural way in which broadband
and angular spread effects can be included in the formalism.

The concept of wave-kinetics has been developed initially to study the
interaction of electromagnetic (EM) waves with plasmas, e.g. in laser
wakefield acceleration, but has meanwhile been extended to other areas of
plasma physics as well~\cite{bing97,silva00,bing96,silva97,tito03}. At its
heart lies the \emph{wave action} $N(t,x,k)$ of the incoming pump wave, which
is defined as $N(t,x,k) = W(t,x,k)/\omega(k)$. Here, $W$ denotes the energy
density of the wave as a function of time, position, and wave number
(i.e. momentum of the wave mode), and $\omega(k)$ is the wave mode frequency
as given by the linear dispersion relation. Since the energy of a wave mode
comes in quanta of size $\hbar\omega(k)$, the quantity $N$ is a measure of the
density of wave quanta or \emph{quasi-particles} for each wave number
$k$. Owing to the fact that wave action is, in general, a conserved
quantity~\cite{whit}, the evolution of $N(t,x,k)$ is then governed by a
Vlasov-type equation, of which the factors $dx/dt$ and $dk/dt$ are provided by
a pair of Hamiltionian equations of motion, also known as the paraxial-ray or
ray-tracing equations. The specific approach allows one to follow the
propagation of a broad spectrum of wave modes simultaneously, as opposed to
conventional approaches that allow only a limited number of modes to be
followed in a single simulation.

A range of complex phenomena can be investigated using the wave-kinetic
approach that cannot be studied by other means. These include: photon
acceleration and deceleration, Landau-like damping of plasma waves by photons,
photon wave breaking and turbulence, broadband modulational instabilities, and
stretching/steepening of short laser pulses. Applications in laser-plasma
interaction are many, such as wakefield excitation, broadband laser-plasma
interaction (e.g. induced spatial incoherence), strong plasma turbulence
(e.g. the Langmuir wave collapse), and so on. There may also be numerous
applications in other areas involving the interaction of a broad spectrum of
fast waves with a plasma.

The code described in this paper solves the Vlasov equation for the wave
action $N(t,x,k)$ using the well-known particle-in-cell (PIC)
method~\cite{dawson83,bird}. There are only two other known attempts to
write such a code: one by Tappert et al.~\cite{tap71}, who did not include
the plasma response to the EM waves in their model, and one by  Silva et
al.~\cite{silva00}, who used a quasi-static model for the plasma response.
To our knowledge, this code is the first to use a fully non-linear,
relativistic, non-quasi-static model for the plasma response.
As the particles involved are (quasi-)photons in our case,
we have decided to call it a photon-in-cell code. This code has been used to
simulate the propagation of a short, intense laser pulse through a slab of
underdense plasma. The particle model for the photons has been coupled to a
cold-fluid model for the plasma. Even though this model is rather basic, the
code produces quite remarkable results. If necessary, the fluid model can be
replaced with more advanced models (PIC or Vlasov) for the plasma
evolution. Some basic features of the code: it is less computationally
demanding than a full PIC code, provides a much simpler description of the EM
waves than the ``slowly evolving envelope'' approximation, gives better
physical insight into the EM field dynamics than most existing codes, and
allows for an easy description of broadband, incoherent light pulses.

This paper is organized as follows. In Section 2, a description of the theory
of wave kinetics, in particular photon kinetics, is given. Section 3 contains
a detailed description of the numerical code we have developed to based on
the wave-kinetic model. Section 4 contains several examples of typical
applications of the code. These serve to demonstrate the capabilities of the
code, as well as the advanced diagnostics that it provides, which give a
deeper insight into the physics underlying the simulation results than can be
obtained with many commonly used simulation techniques. Finally, Section 5
contains the conclusions that can be drawn from the presented results, as well
as an overview of the ways in which the code can be extended and/or improved.

\section{Theoretical background}

The theoretical foundation for the quasi-particle approach consists of three
main parts: the Wigner-Moyal description of classical EM fields and the
associated Vlasov-like equation describing the conservation of
quasi-particles, the dispersion relation for the quasi-particles and the
associated kinetic equations, and the ponderomotive force exerted by the
quasi-particles on the plasma. We will treat these briefly.

The Wigner-Moyal formalism for the description of waves, which is well-known
from quantum-mechanical theory, can also be employed to describe the
propagation of classical EM fields~\cite{tap71b,mend01}. In the framework of
this formalism, the equation describing wave propagation through a dispersive
medium (i.e. a plasma) assumes the familiar Vlasov-like description when the
fields evolve on a much faster time scale than the medium, and quantities like
the susceptibility $\chi$ evolve on the same slow time scale as the medium.
The central quantity in the Wigner-Moyal description is the quasi-particle
density $N(t,x,k)$, given by $N(t,x,k) = W(t,x,k)/\omega(k)$, where $W(t,x,k)$
is the energy density of the Fourier component of the field with wave number
$k$ at time $t$ and position $x$. For a given EM field $\vec{E}(t,x)$, the
corresponding quasi-particle density is obtained using the Wigner function:
\begin{equation}
\label{eq:wigner}
F(t,x,k) = \int E(t,x+s/2) \cdot E^*(t,x-s/2) \exp(-i k s) \mathrm{d}s
\end{equation}
from which $N(t,x,k)$ can be derived as follows:
\begin{equation}
\label{eq:qpden}
N(t,x,k) = \frac{\epsilon_0}{8\hbar} \left( \pd{D}{\omega} \right)_{\omega(k)}
\cdot F(t,x,k).
\end{equation}
Here, $D=0$ is the dispersion relation of the medium in which the
quasi-particles are propagating, and $\omega(k)$ is derived from the same
dispersion relation. From $N$, both the square of the electric field and its
vector potential can be derived quickly:
\[
E^2(t,x) = \frac{\hbar}{\epsilon_0}\int \frac{\mathrm{d}k}{(2\pi)^3}
\omega(k) N(t,x,k), \quad
A^2(t,x) = \frac{\hbar}{\epsilon_0}\int \frac{\mathrm{d}k}{(2\pi)^3}
N(t,x,k)/\omega(k).
\]

The quasi-particle density $N$ obeys the following equation:
\begin{equation}
\label{eq:liou}
\frac{dN}{dt} = \pd{N}{t} + \frac{dx}{dt}\pd{N}{x} + \frac{dk}{dt}\pd{N}{k} =
S(t,x,k,N),
\end{equation}
where the source term $S$ contains slow variations of the refractive index
$\epsilon$ of the medium, as well as terms governing the creation and
absorption of quasi-particles. For a broad range of applications, these
contributions can be neglected, i.e. $S \approx 0$, and (\ref{eq:liou})
reduces to the Liouville equation for the quasi-particle density $N$.

When there is a clear separation between fast and slow time scales, the
propagation of radiation through a medium is mostly governed by the
corresponding linear dispersion relation. For example, for EM radiation
propagating through a plasma, we have $D=\omega^2 - \omega_p^2(x,t)/\gamma -
c^2 k^2 = 0$, where $\omega_p$ denotes the local electron plasma frequency,
and $\gamma$ is the relativistic mass correction factor. From this dispersion
relation, a relation of the form $\omega=\omega(k)$ can be derived,
e.g. $\omega(k) = \sqrt{\omega_p^2(x,t)/\gamma + c^2 k^2}$. Taking into
account that $\omega(k)$ serves as the effective Hamiltonian for the canonical
ray equations, we derive the factors $dx/dt$ and $dk/dt$ in (\ref{eq:liou})
from the well-known ray tracing equations:
\begin{equation}
\label{eq:kin}
\frac{dx}{dt} = -\frac{\partial D/\partial k}{\partial D/\partial\omega} =
\pd{\omega}{k}, \quad \frac{dk}{dt} = 
\frac{\partial D/\partial x}{\partial D/\partial\omega} = -\pd{\omega}{x}.
\quad \frac{d\omega}{dt} =
-\frac{\partial D/\partial t}{\partial D/\partial\omega} = \pd{\omega}{t}
\end{equation}
These equations allow one to describe the full evolution of the quasi-particle
density $N(t,x,k)$ in a plasma of given density. Numerical implementation can
be realized by means of e.g. a Vlasov or a particle-in-cell code; the latter
approach will be explored in this paper.

The third topic we need to address here is the action of the quasi-particles
on the medium.  Together with the evolution equation for $N$, knowledge of
this action will allow us to produce a self-consistent description of the
interaction between the radiation and the medium through which it
propagates. The action is given by the ponderomotive pressure $F_p(x,t)$
\emph{per unit volume}, as found in \cite{silva99}:
\[
F_p(x,t) = g_{sq} \int \frac{\mathrm{d}k}{(2\pi)^3} N(t,x,k) \nabla
H_{\mathrm{eff}}-g_{sq} \int \frac{\mathrm{d}k}{(2\pi)^3} \nabla
\left[ N(t,x,k) \pd{H_{\mathrm{eff}}}{n_{bg}} n_{bg} \right],
\]
where $g_{sq}$ is the quasi-particle statistical weight (accounts for spin
degeneracy of the QPs), $H_{\mathrm{eff}}$ is the effective Hamiltonian for the
canonical ray equations, and $n_{bg}$ is the background density of the
medium. For photons propagating through a plasma with background density
$n_0$, we have $H_{\mathrm{eff}} = \omega(k) = \sqrt{\omega_p^2(x,t)/\gamma
+ c^2 k^2}$, and
\begin{equation}
\label{eq:pond}
F_p(x,t) = -\frac{\omega_{pe}^2}{2\gamma} g_{ph} \nabla \int
\frac{\mathrm{d}k}{(2\pi)^3} N(t,x,k) / \omega(k).
\end{equation}
where $g_{ph}$=1 ($g_{ph}$=2) for circular (linear) polarization, and $n_0$
denotes the background plasma electron density. For a plane EM wave
$(\omega_0, k_0)$, the QP density is given by $N=|E_0|^2/(8g_{ph} \pi \hbar
\omega_0) \delta (k-k_0)$, and the ponderomotive force \emph{acting on a single
electron} takes the more familiar form ($A^2$ denotes the wave's vector
potential):
\[
F_p(x,t) = -\frac{e^2}{2 m_e \gamma} \nabla A^2,
\]
Together with a model for the evolution of the plasma electrons, the equations
(\ref{eq:qpden}), (\ref{eq:liou}), (\ref{eq:kin}) and (\ref{eq:pond}) allow us
to describe the propagation of an arbitrary laser pulse in an underdense
plasma in a self-consistent manner. In the next section, we will describe the
numerical implementation of these equations that we have developed.

\section{The photon-in-cell algorithm}

\subsection{Description of the algorithm}

The photon-in-cell algorithm is based on the fact that the equation for the
evolution of the quasi-particle density (\ref{eq:liou}) is nothing different
from the well-known Vlasov equation for the evolution of the plasma electron
or ion density. Thus, a number of proven techniques are available to us for
the numerical study of that equation. We have chosen to employ the so-called
particle-in-cell method. This method makes full use of the fact that the
quasi-particle number density $N$ is conserved along the flow lines in
quasi-particle phase space. In other words, if $N$ is the superposition of a
certain number of $\delta$-distributions at $t=0$, i.e.  $N(0,x,k) = \sum_i
\delta(x-x_i(0)) \delta(k-k_i(0))$, then the Vlasov equation (\ref{eq:liou})
guarantees that $N$ will remain the superposition of the same number of
$\delta$-distributions for all time, and one only has to keep track of the
locations $(x_i(t),k_i(t))$ of the distributions in phase space. This
evolution is governed by the characteristics of the Vlasov equation:
\[
\frac{dx_i}{dt} = \pd{\omega}{k}(t,x_i(t),k_i(t)), \qquad \frac{dk_i}{dt} =
-\pd{\omega}{x}(t,x_i(t),k_i(t)),
\]
where $\omega(t,x,k)$ is provided by the dispersion relation for the
quasi-particles under consideration. (See below.)

For the plasma electrons, we have employed a 1-dimensional cold fluid
model. The model is fully non-linear and relativistic, as linear and weakly
non-linear models yielded unsatisfactory results. The ions are treated as a
static background. First we apply the following scaling to our equations:
$t \rightarrow \omega_p t$, $x \rightarrow \omega_p x/c$, $\omega \rightarrow
\omega/\omega_p$, $k \rightarrow ck/\omega_p$, $p \rightarrow p/(m_e c)$,
$n \rightarrow n/n_0$, $E \rightarrow eE/(m_e \omega_p c)$, $A \rightarrow
eA/(m_e c)$. Here, $n_0$ is the plasma background density, $\omega_p = (e^2
n_0/(\epsilon_0 m_e) )^(1/2)$ denotes the corresponding plasma frequency, and
other symbols have their usual meaning. The scaled equations then read:
\begin{eqnarray}
\label{eq:dndt}
\pd{n}{t} + \pd{}{x}\left(\frac{np}{\gamma}\right) &=& 0,\\
\pd{p}{t} + \frac{p}{\gamma} \pd{p}{x} &=& -E -\frac{1}{2\gamma} \pd{A^2}{x},\\
A^2 &=& \int \frac{\mathrm{d}k}{(2\pi)^3} \frac{N(t,x,k) }{\omega(k)},\\
\pd{E}{x} &=& 1-n.
\end{eqnarray}
For a circularly polarized laser pulse, we have $\gamma = \sqrt{1+p^2 + A^2}$,
and the evolution equation for $p$ simplifies to
\begin{equation}
\label{eq:dpdt}
\pd{p}{t} + \pd{\gamma}{x} = -E.
\end{equation}
For the scaled dispersion relation, we now find
\[
\omega(t,x,k) = \sqrt{n/\sqrt{1+p^2+a^2} + k^2},
\]
from which we find the equations of motion for the photons:
\begin{equation}
\label{eq:dxdt}
\frac{dx_i}{dt} = \frac{k_i}{\omega(t,x_i,k_i)}, \qquad \frac{dk_i}{dt} =
-\frac{1}{2\omega(t,x_i,k_i)} \pd{}{x}\left(\frac{n}{\gamma}\right).
\end{equation}
The equations (\ref{eq:dndt}), (\ref{eq:dpdt}) and (\ref{eq:dxdt}), together
with the equations for $E$, $A^2$, $\gamma$ and $\omega$, form a closed system
for the evolution of the combined radiation-plasma system. We have created a
numerical implementation of this scheme based on the following steps:
\begin{enumerate}
\item Create a grid spanning the simulation box; quantities like $n$, $p$, and
$A^2$ will be known on this grid;
\item For a given laser pulse, initialize the photon number density using
(\ref{eq:wigner}) and (\ref{eq:qpden});
\item Load the quasi-particles in such a way that their distribution matches
the number density calculated in the previous step; project them onto the grid
to obtain $A^2$;
\item Initialize $n$, $p$, and other field quantities;
\item Advance $n$ and $p$;
\item Advance $x_i$ and $k_i$ for all quasi-particles; 
\item Advance all other field quantities;
\item Project the photons onto the grid to obtain the new value of $A^2$.
\end{enumerate}
The steps 5-8 are repeated for each time cycle.

There are a number of practical issues arising during the implementation of
the above scheme.  These will be discussed in the next subsection.

\subsection{Implementation}

The practical implementation of the above scheme consists of three distinct
parts: initialization, field and particle advance, and particle
projection. Each will be treated here briefly.

{\bf Initialization.} The initialization stage works as follows. First, a
monospaced grid is created. On this grid, the plasma electron density is
initialized to the (prescribed) value of the background ion density, while the
plasma fluid momentum is initialized to zero, so the plasma is initially
neutral and stationary. Then for each grid cell, the (prescribed)
quasi-particle density is integrated over all $k$ values to determine the
total number of quasi-particles for that cell. Next the quasi-particles are
loaded for each cell; the positions are allocated deterministically, while the
wave numbers are allocated randomly. This way, the effective quasi-particle
density corresponds to an approximation of the prescribed one that is
piecewise linear in the $x$-direction and piecewise constant in the
$k$-direction. Once they are loaded, the particles are projected onto the grid
to determine $A^2$.  This is not a straightforward procedure: to calculate
$A^2$ one needs $\omega(k_i)$ for each particle, to calculate $\omega(k_i)$
one needs the Lorentz factor $\gamma(x_i)$, and to calculate $\gamma(x_i)$ one
needs $A^2$ again. To overcome this problem, we set $\gamma=1$ everywhere, and
then alternatingly project the photons and recalculate $\gamma$. Eventually
this process leads to a converging value for $A^2$; for our case, it was
sufficient to repeat the procedure twice.

{\bf Field and particle advance.} For the advancement of both fields and
particles, we resort to a second-order Runge-Kutta scheme. While this is a
straightforward procedure for the canonical quasi-particle equations, the
applicability of this scheme to the fluid equations is not immediately
apparent. We recall that the 1-D fluid equations read:
\[
\pd{n}{t} = -\pd{}{x}\left(\frac{np}{\gamma}\right), \qquad \pd{p}{t} =
-E-\pd{\gamma}{x}.
\]
By discretizing the right-hand side of these equations, to get rid of the
$x$-derivatives, we get:
\begin{equation}
\label{eq:discr}
\begin{array}{r l}
\displaystyle \pd{n}{t} &= \displaystyle \frac{-1}{2h}\left( (np/\gamma)(x+h) 
- (np/\gamma)(x-h)\right),\\
\\
\displaystyle \pd{p}{t} &= \displaystyle \frac{-1}{2h}\left( E(x+h)-E(x-h)
+\gamma(x+h) - \gamma(x-h) \right).
\end{array}
\end{equation}
If we then advance $n$ and $p$ across the whole grid at once, instead of
node-by-node, the equations (\ref{eq:discr}) can be treated as ordinary
differential equations, to which the Runge-Kutta scheme can be applied as
usual.

In order to properly implement the Runge-Kutta scheme, we have to keep track
of a number of auxiliary quantities; apart from $n$ and $p$, intermediate
values for $E$, $\gamma$, $n/\gamma$, $\partial/\partial x (n/\gamma)$, and
$np/\gamma$ are used.

The Runge-Kutta scheme roughly works as follows:
\begin{enumerate}
\item Advance $n$, $p$, $x$, $k$ for $0.5*\delta t$;
\item Update $\gamma$ using intermediate values for $p$ but old values for
$A^2$;
\item Project photons to obtain intermediate value of $A^2$
\item Update $\gamma$ and all other derived field quantities;
\item Advance $n$, $p$, $x$, $k$ for $\delta t$, using the intermediate values
for all quantities calculated before;
\item Update $\gamma$ using new values for $p$ but intermediate values for
$A^2$;
\item Project photons to obtain new value of $A^2$
\item Update $\gamma$ and all other derived field quantities;
\end{enumerate}
Although this scheme is not the most efficient in terms of memory use, its
benefits are clear. The scheme is very stable and tolerates solutions that are
strongly non-linear. Peak values of $n/n_0=10$ have been reached, i.e. very
close to wave breaking, before the simulation would become unstable and break
down.

{\bf Projection.} We have used quadratic splines to accumulate the
contributions of the particles to $A^2$ onto the grid, as well as to
interpolate field quantities, which are only known on grid nodes, at the
positions of the particles. Although a quadratic spline is more expensive
computationally than a linear one, the use of quadratic splines leads to
significant noise reduction and solutions that are more stable, especially for
long runs. The following spline function has been used:
\[
f(x) = \left\{
\begin{array}{l l}
0.75-x^2, & |x| < 0.5,\\
0.5*(1.5-|x|)^2, & 0.5 \leq |x| < 1.5,\\
0, & |x| \geq 1.5.\\
\end{array}
\right.
\]
This function is continuous with a continuous first derivative, and satisfies
$f(x-1) + f(x) + f(x+1) = 1$ for $|x| < 0.5$. The latter property ensures
proper photon number density conservation.

After the photons have been projected and $A^2$ has been obtained, a four-pass
binomial filter is applied to the result. This has been found to be necessary
to suppress numerical instabilities that otherwise grow uncontrollably and
destroy the result in the long run. These instabilities result from
modulational instabilities (which do not have a well-determined wave length,
unlike parametric instabilities) are growing from any kind of numerical noise
around. Fortunately, the instabilities occur at (much) shorter wave lengths
than the interesting physics, which allows one to filter them out as long as
the grid is not too coarse. The use of a compensation filter turned out to be
undesirable, as it would undo the effect of the filter on the instabilities.

It should be noted that additional filtering cannot be used as a substitute
for the use of quadratic splines. This is due to the fact that filtering has
shown to be very effective on fast-growing instabilities, but less so on
slow-growing instabilities that might show up during (very) long runs.  The
latter kind is much more effectively suppressed by using quadratic splines
instead of linear ones. 

\section{Applications}

To demonstrate the possibilities of the photon-kinetic approach, and to
benchmark the code, we have used it to simulate the interaction of laser
pulses having various lengths with a slab of cold, underdense plasma. The
pulses have a Gaussian-shaped envelope, and as they are taken to be coherent,
their spectral bandwidth is inversely proportional to their length. Their
peak intensity is given by $a_0 = 0.7$.

The plasma slab has a constant background density, which is scaled to 1. The
degree to which the plasma (plasma frequency $\omega_p$) is underdense with
respect to the laser pulse (mean wave number $k_0$) is fully determined by the
parameter $ck_0/\omega_p$. We will therefore characterize our simulations by
the value of this parameter rather than individual values of $\omega_p$ and
$k_0$.

We shall treat two different cases: a ``long'' pulse (having an FWHM of
several times $c/\omega_p$) and a ``short'' pulse (having an FWHM that is less
than $c/\omega_p$). The observed interaction is quite different from one case
to the next, and we will show that the photon kinetic approach is well suited
to explore these differences. We will also demonstrate the wide range of new
diagnostics that come with this approach.

\subsection{Long pulse}

The interaction of a long pulse (FWHM $\approx 24 c/\omega_p$, i.e. some 4
plasma wavelengths) with an underdense plasma is dominated by the modulational
instability. This is caused by the fact that $\nabla A^2$ is too small
initially to drive a significant wakefield. However, the nature of the photon
equations of motion (\ref{eq:dxdt}) is such, that photons tend to clump
together at slight density depressions of the plasma fluid. Conversely, the
ponderomotive force of a clump of photons tends to push the plasma away, so
wherever the photon density is slightly raised, the plasma density will be
depressed, and vice versa. Even the smallest ripples in either the
plasma density or the pulse envelope will be amplified and develop into a
plasma wave with frequency $\sim\omega_p$ and a speed comparable to that of
the driving laser pulse, i.e. $\sim(1-(1-(ck/\omega_p)^2)^{-1})^{1/2}$.

\begin{figure}
\includegraphics[width=0.75\textwidth]{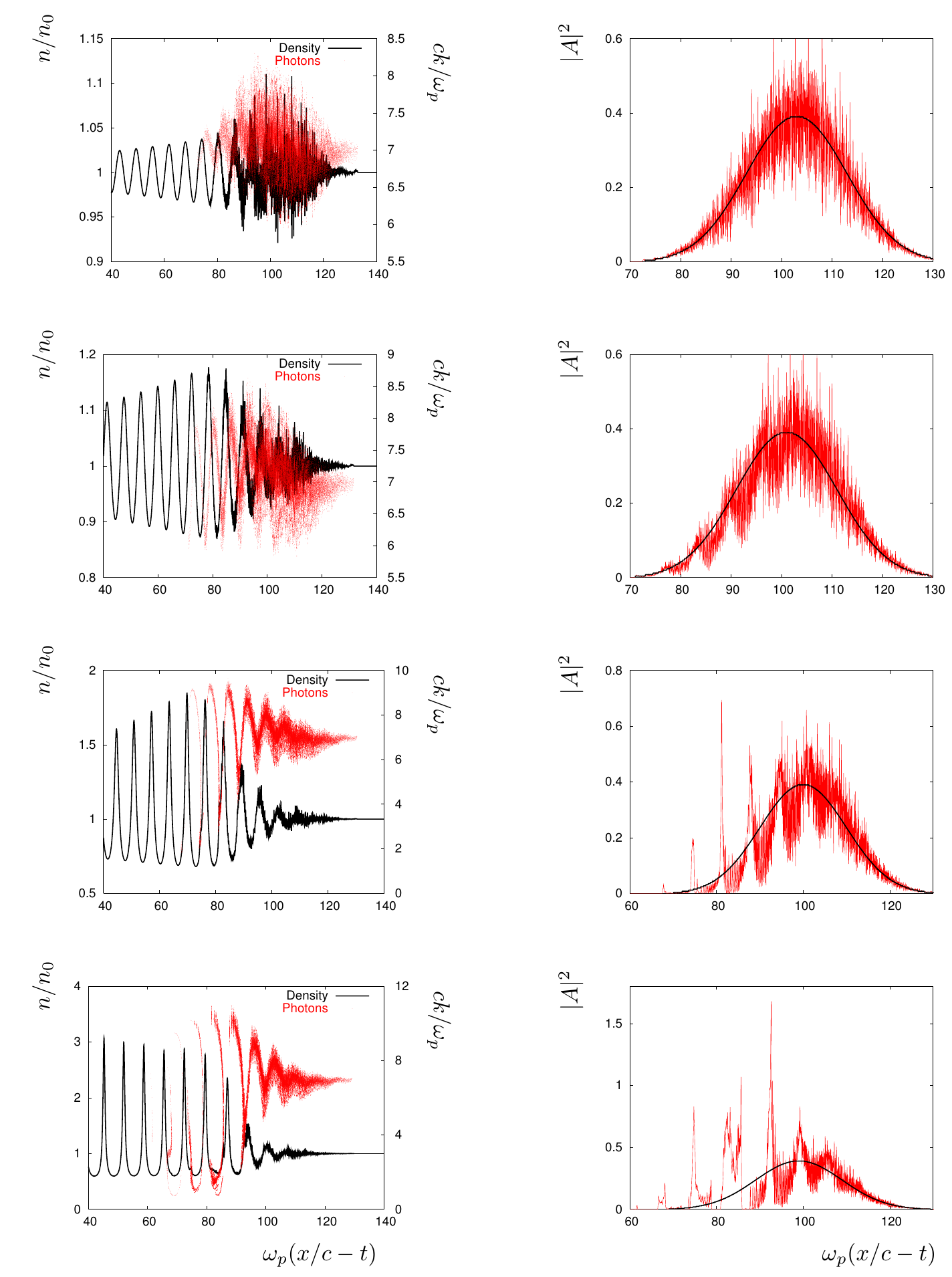}
\caption{Simulation results for the long pulse (FWHM $\approx 24*c/\omega_p$)
  at (a) $\omega_p t=250$, (b) $\omega_p t=375$, (c) $\omega_p t=500$, and
  (d) $\omega_p t=625$. The graph on the left shows $n/n_0$ (black curve, left
  scale) and photon momentum $ck/\omega_p$ (red dots, right scale) versus
  position $\omega_p(x/c-t)$. The growth of the modulational instability, as
  well as photon confinement and turbulence, are clearly present. The graph on
  the right shows $|A|^2$ for the pulse (red curve). The peaks at the back of
  the pulse give evidence of the modulational instability. The black curve
  depicts the original pulse profile for reference.}
\label{fig:1}
\end{figure}

A simulation of a long pulse traversing the plasma has been conducted, the
results of which are displayed in Figure~\ref{fig:1}. The pulse has
an FWHM of about $24 c/\omega_p$,
and a mean wave number $k_0 = 7.0*\omega_p/c$. The spread in wave numbers is
Fourier-matched to the pulse length, in order to have a coherent
pulse. Snapshots of photon $(x,k)$-space, the plasma electron density, as well
as the envelope of the pulse's vector potential, taken at various times, are
displayed.  There is a vast spectrum of interesting phenomena that can be
observed here. First of all, there is photon acceleration/deceleration, which
comes in two flavours: change of $k$ for constant $\omega$, and change of
$\omega$ for constant $k$. The first case corresponds to a photon moving
through a plasma of which the density is constant in time, and corresponds to
optical refraction in a dispersive medium. The second case corresponds to a
photon moving through a plasma of which the density is (locally) constant in
space, but not in time. This type of interaction leads to energy transfer
between the photons and the plasma, and also to a change in colour of the
photons.  Initially, there is mostly energy transfer from the photons to the
plasma, for the creation of the plasma wave. However, energy may also flow in
the opposite direction, in which case the photons are gaining energy and doing
a blue shift at the expense of the plasma wave.

When there is simultaneous acceleration and deceleration of particles, there
is always a good chance of exciting turbulence. This is no different for
quasi-particles. Quasi-particle turbulence works as follows. In the case of an
unchirped laser pulse, the mean value of $k$ is constant through the pulse
initially. When the modulational instability is triggered, there will not only
be a modulation on the spatial distribution of the quasi-particles, but also
on their mean $k$-value, and thus on their (group) velocity. When the
simulation progresses, this modulation increases in amplitude, until such time
that fast quasi-particles from the back of the pulse start to overtake slow
quasi-particles from the front. This actually constitutes ``wave breaking'' of
the quasi-particle wave. Meanwhile, a significant plasma wave has been
created, and most quasi-particles will be trapped in the troughs of this
wave. Fast particles moving to the front of a trough will be decelerated by
the increasing plasma density, move to the back of the trough, will be
accelerated by the increasing plasma density they meet there, and move to the
front again. In quasi-particle $(x,k)$-space (phase space), one observes
groups of quasi-particles, each group confined to its own trough, revolving
around an O-point.

By this time, the plasma density fluctuations are driven far into the
non-linear regime, even if one started out with a pulse intensity that is
still in the linear regime. The reason for this is, that the pulse envelope
has been modulated on the length scale of the plasma wave, and thus exhibits
peaks spaced $c/\omega_p$ apart. This can clearly be seen in the plots on the
right of Figure~\ref{fig:1}. Note that the red curve follows the actual
profile of $|A|^2$, while the black curve denotes the original profile for
reference.  The peaks in $|A|^2$ lead to large values for $\nabla A^2$,
i.e. for the associated ponderomotive force, far in excess of the values
obtained from the shallow slopes of the original envelope. As they are also
spaced approximately one plasma wavelength apart, they act in resonance with
the plasma wave itself, and drive it much more efficiently and to much larger
amplitudes than an unmodulated pulse ever would. When a wakefield is excited
by a long pulse through the mechanism we just described, it is commonly called
\emph{self-modulated} laser wakefield excitation~\cite{esa96}.

\begin{figure}
\includegraphics[width=0.75\textwidth]{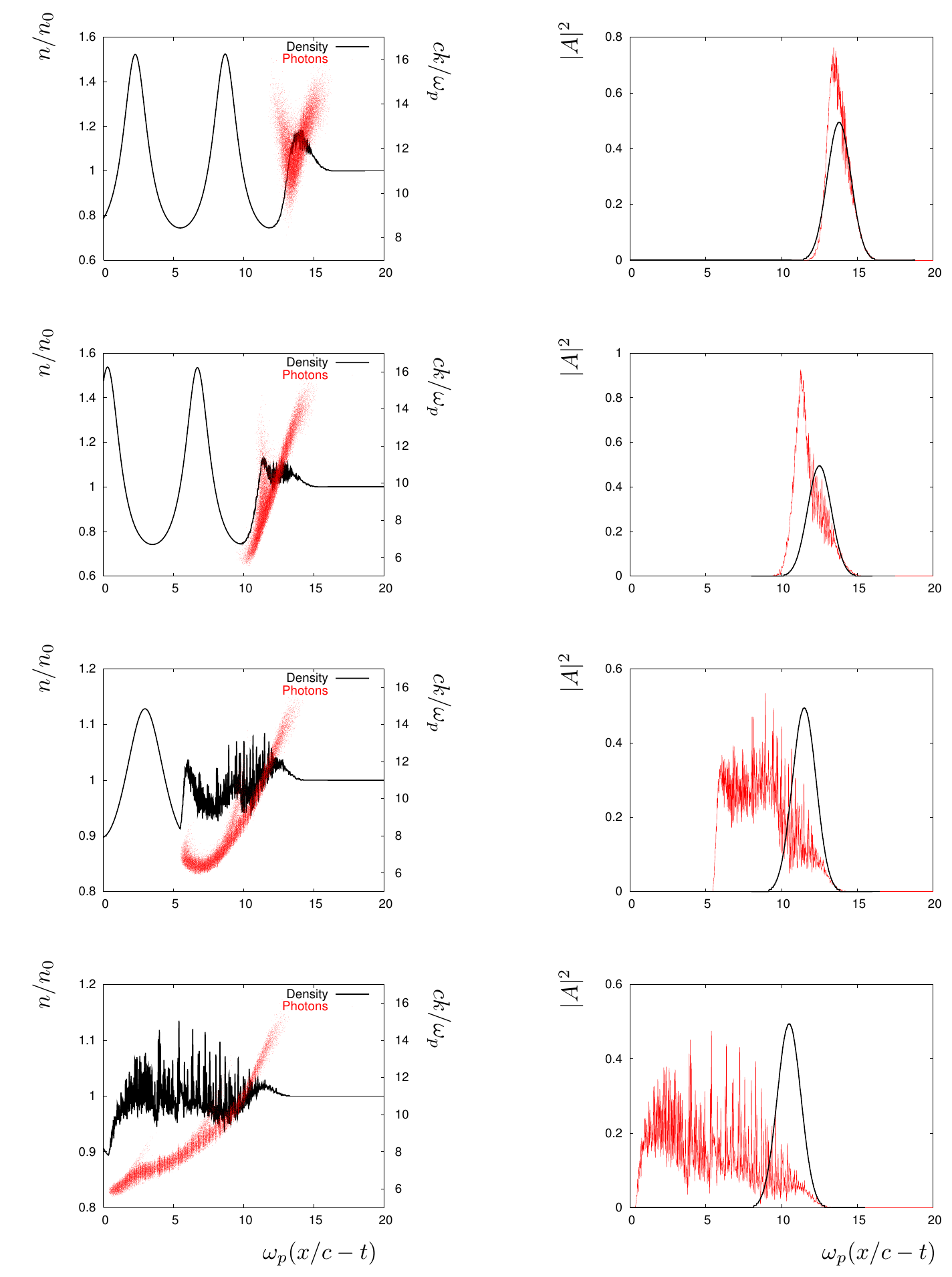}
\caption{Simulation results for the short pulse (FWHM $\approx 1.8*c/\omega_p$)
  at (a) $\omega_p t=467$, (b) $\omega_p t=933$, (c) $\omega_p t=1400$, and
  (d) $\omega_p t=1867$. The graph on the left shows $n/n_0$ (black curve, left
  scale) and photon momentum $ck/\omega_p$ (red dots, right scale) versus
  position $\omega_p(x/c-t)$. It is obvious that photons in the middle of the
  pulse lose much more energy than those at the ends, resulting in stretching
  and chirping of the pulse. The graph on the right shows $|A|^2$ for the
  pulse (red curve). One observes initial pulse compression and steepening,
  whereas the pulse is stretched and flattened at later times. The black curve
  depicts the original pulse profile for reference.}
\label{fig:2}
\end{figure}

\subsection{Short pulse}

The interaction of a short pulse (FWHM $\approx 1.8*c/\omega_p$, i.e. about
0.3 plasma wavelength) with the plasma is dominated by photon deceleration and
the transfer of energy from the photons to the plasma. For the case of a
reasonably underdense plasma, the pulse velocity will be close to $c$, and the
wave buckets will be approximately $c/\omega_p$ in length, i.e. longer than
the pulse itself. For this reason, the photons will mostly be grouped at the
front of the first wave bucket, and there will be neither photon acceleration
nor turbulence. Also, the modulational instability is not expected to occur.

A simulation has been conducted, the results of which are displayed in
Figure~\ref{fig:2}. The pulse has 
an FWHM of about $1.8 c/\omega_p$, and a mean wave number $k_0 =
14.0*\omega_p/c$. Again, the spread in wave numbers is Fourier-matched to the
pulse length. In the figure, as in Figure~\ref{fig:1}, graphs of the photon
phase space and $n/n_0$ are shown on the left, while the evolution of the
pulse's envelope is shown on the right. At the start of the interaction, we
observe that the pulse excites a wakefield, and that the photons lose energy
and momentum in the process. It is interesting to note that the photons right
after the centre of the pulse lose the most energy. This follows from the fact
that $d\omega/dt = \partial\omega /\partial t = (1/2\omega) \partial
n/\partial t$, and $\partial n/\partial t$ is most negative direcly behind the
pulse centre. The photons at the the pulse's extremities find themselves in
regions where $|\partial n/\partial t|$ is small, and hardly see any energy
change. This leads to a characteristic V-shape of the photons in phase space.

Looking at the evolution of the pulse envelope, i.e. $|A|^2$, we find that it
initially steepens while its peak value increases. This is a direct
consequence of the decrease of $\omega$ for the particles at the centre of the
pulse. As $|A_k|^2 = N_k/\omega_k$, and $N_k$ is conserved, any decrease in
$\omega_k$ will initially lead to an increase in $|A|^2$. Only after the
photons are spread out along a larger length, will the peak amplitude of the
vector potential decrease. See also Tsung \emph{et al.}~\cite{tsung} for a
detailed description of this effect.

As the interaction progresses, the slowest photons (low $k$-value) will start
to lag behind with respect to the fastest photons. This will cause the pulse
to be stretched and chirped (i.e. tilted in phase space). Due to the continued
energy transfer from the photons to the plasma, the mean values of $\omega$
and $k$ for the pulse as a whole will decrease, causing it to undergo a
general redshift. We also find that the pulse (moving with $v_g < c$) is
lagging behind with respect to the simulation box (moving with $c$), and more
so for decreasing photon momentum.

At even later times, a clump of photons that were at the centre of the pulse
originally, has lost so much energy and is lagging so far behind, that they
now form the back of the pulse. These photons are now encountering a region of
increasing plasma density at the back of the first wave bucket, causing their
momentum and energy to increase once more, at the expense of the wakefield.
This might eventually lead to photon turbulence, as in the case of the long
pulse with modulational instability, if it weren't for the fact that the
pulse, already having spent most of its energy, is no longer able to generate
a wakefield of sufficient intensity for this. Eventually, these photons will
escape from the first wave bucket, leading to further stretching of the pulse
and depletion of its energy.

This particular simulation provides new insights into the internal
dynamics of a short, intense laser pulse driving a wakefield. It shows us
which part of the pulse is driving the wake, and also provides a more
thorough picture into the chirping, stretching and flattening of the pulse
that occurs when the pulse loses a significant amount of its energy to the
wakefield. Such knowledge can be used, for example, to tailor laser pulses in
order to increase the efficiency with which they drive the wakefield.

\section{Conclusions}

In this paper, we have investigated the wave kinetic approach as a novel way
to study non-linear wave-plasma interactions. We have shown that this approach
is particularly suitable in the case of a random phase pump wave interacting
with coherent structures in the plasma, while its extended set of available
diagnostics provides a deeper insight into the turbulent dynamics of the fast
wave modes.

We have developed and tested a PIC-type code to implement the photon-kinetic
model for the interaction of EM waves with an underdense plasma. In doing so,
we have identified the important issues in the development and operation of
such a code. First of all, one needs to choose a proper model for the plasma
response. The linear model we tried first proved to be much too limited for
the applications we had in mind, and introduced additional noise into the
simulation as well. The non-linear relativistic model we are using today
serves well for a range of applications. Second, the proper projection of the
particles onto the grid. Even though it may be tempting to use a linear spline
because it is fast and easy to use, our results show that a quadratic spline
performs better especially for longer runs, and is to be preferred. Finally,
there is the issue of filtering. As with most particle codes, our code suffers
from noise; it has however been found that the growth of noise can be kept
under control by proper filtering of the (square of the) vector
potential. However, this does not lead to a complete obliteration of the
noise, and we may have to resort to other measures to achieve that goal.

The code has been tested on the interaction of laser pulses with a slab of
underdense plasma. The pulse has been represented by quasi-photons, while a
fluid model has been employed for the plasma. Two cases have been
investigated: a ``long'' pulse (several plasma periods in lenght) and a
``short'' pulse (shorter than a single plasma period). For the long pulse, we
have observed the following phenomena: photon acceleration/deceleration,
photon ``wave breaking,'' photon turbulence, and modulational instabilities.
For the short pulse we have found the following phenomena: initial pulse
compression and steepening, selective energy transfer to the plasma wave
(photons at the centre lose most, photons at both ends hardly affected), and
pulse chirping, stretching and flattening at later times.

The photon kinetic treatment offers new diagnostics that provide a deeper
insight into the dynamics of wave-plasma interaction than available with
present-day codes that treat the EM waves as coherent fields. The simulation
results reveal a complex interaction structure that remains hidden
otherwise. This is one of the strong points of the wave kinetic approach.

The deeper insight into pulse dynamics can be employed to tailor pulses
to produce a wakefield with some desired properties more effectively. As
systems to deliver tailor-made laser pulses are becoming more and more
available, numerical simulations to accompany the experimental results will
soon be badly needed. The photon kinetic method can supply the demand.

There is a number of issues that deserve further investigation. First of all,
the code should be generalized to two and three dimensions. This is a
necessary step, as most other codes in the field employ at least two spatial
dimensions. Also, given the speed of current computers and the expected
improvements, there is no reason not to do this. Second, we need better ways
to deal with the noise responsible for short-wavelength modulational
instabilities should be found. We can fight the symptoms using filtering, but
digging out the root cause is the better option in the long run, especially
since we may wish to apply the wave-kinetic approach to other types of
wave-plasma interaction. Third, we wish to extend the the wave-kinetic
approach, and in particular its numerical implementation, to other types of
fast wave-slow wave interactions. Restricting ourselves to photons will not do
the method justice. There are already many proven codes in the field to deal
with the problem of laser-plasma interaction. Whereas this renders
benchmarking of our code easier, it also means that there is not much
territory left to cover in that direction. However, there exist many other
turbulent wave-plasma interactions for which no suitable numerical model has
been developed so far, such as interactions between fast and slow plasma waves
in a magnetized plasma. Such waves occur in a vast range of different
situations, but no numerical methods to attack them efficiently have been
developed to date. However, now that we have demonstrated the wave-kinetic
method successfully, its application to a wide spectrum of turbulent
wave-plasma interactions lies within reach.


\begin{thebibliography}{99}

\bibitem{sagdeev} Sagdeev, R. Z. and Galeev, A. A., \emph{Nonlinear Plasma
Theory\/} (Benjamin, New York, 1969).
\bibitem{bes73} Besieris, I.M. and Tappert, F.D., J. Math. Phys.{\bf 14}, 704, 
(1973).
\bibitem{wig32} Wigner, E., Phys. Rev. {\bf 40}, 749 (1932).
\bibitem{moyal49} Moyal, J.E., Proc. Cmab. Phil. Soc. {\bf 45}, 99, (1949).
\bibitem{peier55} Peierls, R.E., \emph{Quqntum Theory of Solids\/} (Oxford
  University Press, Oxford, 1995).
\bibitem{mcdon88} McDonald, S.W., Phys. Rep. {\bf 158}, 337, (1988).
\bibitem{kadom65} Kadomtsev, B.B., \emph{Plasma Turbulence\/} (Academic Press,
  London, 1965).
\bibitem{bing97} R. Bingham \emph{et al.}, Phys. Rev. Lett. {\bf 78}, 247
(1997).
\bibitem{silva00} L.O. Silva \emph{et al.}, IEEE Trans. Plas. Sci. {\bf 28},
1202 (2000).
\bibitem{dawson83} J. M. Dawson, Rev. Mod. Phys. {\bf 55}, 403-447 (1983).
\bibitem{bird} C. K. Birdsall and A. B. Langdon, \emph{Plasma Physics via
Computer Simulation\/} (Institute of Physics Publishing, Bristol and
  Philadelphia, 1991).
\bibitem{bing96} R. Bingham \emph{et al.}, Physics Letters A {\bf 220},
107 (1996).
\bibitem{silva97} L. O. Silva and J. T. Mendon\c{c}a, Phys. Rev. E {\bf 57},
3423 (1998).
\bibitem{tito03} J. T. Mendon\c{c}a, R. Bingham, P. K. Shukla, Phys. Rev. E
{\bf 68}, 0164406 (2003).
\bibitem{whit} G. B. Whitham, \emph{Linear and Nonlinear Waves\/}
(John Wiley \& Sons, New York, 1974).
\bibitem{tap71} F. D. Tappert, W. J. Cole, R. H. Hardin and N. J. Zabusky,
in {\em Proceedings of the Fourth Conference on Numerical Simulation of
  Plasmas, 1970, Washington, D.C.,} edited by J. Boris and R. Shanny
(Office of Naval Research, Arlington, VA, 1971), 196.
\bibitem{tap71b} F. D. Tappert, SIAM Rev. (Chronicle), {\bf 13}, 281 (1971).
\bibitem{mend01} J. T. Mendon\c{c}a, {\em Theory of Photon Acceleration,\/}
Series in Plasma Physics (Institute of Physics Publishing, Bristol and
Philadelphia, 2001).
\bibitem{silva99} L. O. Silva, R. Bingham, J. M. Dawson, and W. B. Mori,
Phys. Rev. E {\bf 59}, 2273 (1999).
\bibitem{esa96} E. Esarey, P. Sprangle, J. Krall, and A. Ting, IEEE Trans.
Plas. Sci. {\bf 24}, 252 (1996), and references therein.
\bibitem{tsung} Tsung \emph{et al.,} Proc. Nat. Ac. Sci. USA {\bf 99} (1),
  29-32 (2002).

\end{thebibliography}
\end{document}